\documentclass[aps,prb,twocolumn,preprintnumbers]{revtex4}
\usepackage{bm,hyperref}
\usepackage{amsmath}
\pdfoutput=1
\usepackage{graphicx}
\usepackage{soul}
\usepackage{amsfonts}
\usepackage{amssymb}
\usepackage{dcolumn}
\usepackage{amsmath}
\usepackage{lipsum}  
\usepackage{tikz, pgfplots}
 \usetikzlibrary{plotmarks}
  \usetikzlibrary{arrows.meta}
  \usepgfplotslibrary{patchplots}
  \usepackage{grffile}
  \newlength\figureheight
  \newlength\figurewidth

\newcommand{\mv}[1]{\ensuremath{{\bm #1}}}

\newcommand{\curl}{\mv{\nabla}\times}

\begin{document}
\title{Beyond local effective material properties for metamaterials}

\author{K. Mnasri{$^{*1,2}$}, A. Khrabustovskyi{$^{3}$}, C. Stohrer{$^{2}$},\\  M. Plum$^{3}$, and C. Rockstuhl{$^{1,4}$}}
\address{
{$^{1}$}Institute of Theoretical Solid State Physics, Karlsruhe Institute of Technology, Karlsruhe, Germany\\
{$^{2}$}Institute for Applied and Numerical Mathematics, Karlsruher Institut f\"ur Technologie,
Englerstrasse 2, 76128, Karlsruhe, Germany\\
{$^{3}$}Institute for Analysis, Department of Mathematics, Karlsruhe Institute of Technology, Englerstrasse 2, 76131 Karlsruhe, Germany \\
{$^{4}$}Institute of Nanotechnology, Karlsruhe Institute of Technology, Karlsruhe, Germany\\
$^*$Corresponding author: karim.mnasri@kit.edu
}

\begin{abstract}
To discuss the properties of metamaterials on physical grounds and to consider them in applications, effective material parameters are usually introduced and assigned to a given metamaterial. In most cases, only weak spatial dispersion is considered. It allows to assign local material properties, i.e.~a permittivity and a permeability. However, this turned out to be insufficient. To solve this problem, we study here the effective properties of metamaterials with constitutive relations beyond a local response and take strong spatial dispersion into account. The isofrequency surfaces of the dispersion relation are investigated and compared to those of an actual metamaterial. The significant improvement provides evidence for the necessity to use nonlocal material laws in the effective description of metamaterials. The general formulation we choose here renders our approach applicable to a wide class of metamaterials.
\end{abstract}

\maketitle
\section{Introduction}
Assigning effective material properties is a main problem in the macroscopic description of optical metamaterials \cite{PhysRevB.84.075153, PhysRevE.81.036605}. The desire of replacing such materials by hypothetical, homogeneous ones rises from the simplification when discussing them within a physical framework. Once a metamaterial is homogenized, it might be described and discussed on the same level as a natural material by its effective material properties. Moreover, addressing numerically the wave propagation inside a homogeneous material is more efficient, i.e.~less computational effort is required, than in performing a rigorous computation of the full structure. However, it is of utmost importance that both descriptions for the same metamaterial, i.e.~the actual and the homogenized metamaterial, provide the same response, up to a certain accuracy, to the electromagnetic field. Otherwise the homogenization procedure is meaningless.

Various techniques have been established to assign effective material properties to optical metamaterials \cite{PhysRevB.86.035127,PhysRevE.85.066603,Sihvola2013364,PhysRevB.75.115104,PhysRevB.87.235136,Chipouline201277,PhysRevB.88.125131,PhysRevB.93.024418,Grahn:13,1367-2630-15-11-113044,Shevchenko:17}. However, most previous techniques assume local constitutive relations, i.e.~in a homogeneous material, the functional dependency of the auxiliary fields, $\mv{D}$ and $\mv{H}$, is a linear combination of the macroscopic fields $\mv{E}$ and $\mv{B}$, in which the coefficients are spatially independent \cite{WSEAS9vol4sep2007,0966814398}
\begin{align}
\mv D(\mv r, \omega) &= \hat{\epsilon}(\omega) \mv E(\mv r,\omega) +\hat{\xi}(\omega)\mv B(\mv r,\omega)\,, \nonumber  \\
\mv H(\mv r, \omega) &= \hat{\mu}^{-1}(\omega) \mv B(\mv r,\omega) +\hat{\zeta}(\omega)\mv E(\mv r,\omega) \,. \label{eq:LocalBinaisotropy}
\end{align}
These equations are usually called local bi-anisotropic material laws, with effective material parameters $\hat{\epsilon}$, $\hat{\mu}$, $\hat{\xi}$ and $\hat{\zeta}$, where the hats refer to the quantities being tensors.
 A comprehensive review might be found in Ref. \onlinecite{9056993275}.
Although such constitutive relations are usually assumed, they exhibit several limitations \cite{PhysRevB.81.035320,PhysRevLett.115.177402,0957-4484-26-18-184001}. 
In particular, the effective properties are only tensors that do not depend on the considered spatial frequency. However, they turn out to be inadequate when considering light propagation inside the metamaterial in an arbitrary direction and not just in the direction that was considered in the retrieval procedure \cite{PhysRevB.77.195328,PhysRevB.81.035320}. Moreover, whereas it can be safely expected that such constitutive relations are valid when considering metamaterials for which the operational wavelength is much longer than the size of the unit cell, they fail to be predictive for most metamaterials that are operated in a regime where the wavelength is not much smaller than size of the unit cell but only smaller. Such operational regime, unfortunately, is necessary to observe many relevant dispersive effects.

To overcome these limitations, we propose in this work two formulations for advanced constitutive relations in order to model metamaterials beyond a local response and to take strong spatial dispersion into account. The applicability of these models is investigated. The justification for such models is derived by studying here the isofrequency surfaces of the dispersion relation obtained by our models that are compared to those of an actual metamaterial when considering the fundamental mode, i.e.~the mode with the smallest imaginary part in the propagation constant. We find clear evidence that it is necessary to consider strong spatial dispersion, i.e.~nonlocal constitutive relations in the effective description of the metamaterials. We emphasise that in this contribution we study the bulk properties of the metamaterials in terms of the dispersion relation and/or isofrequency plots for the eigenmodes that are supported by the metamaterial. To make that approach viable requires also the formulation  of additional boundary conditions to solve the problem how the modes couple at the interface from one medium to the other. These additional boundary conditions for the materials considered here are beyond the scope of this contribution and will be discussed elsewhere\footnote{A. Khrabustovskyi, K. Mnasri, M. Plum, C. Rockstuhl, C. Stohrer, ``in preparation''}. In the context of nonlocal constitutive relations, comparable approaches were already employed to effectively homogenize metamaterials beyond their local regime
\cite{acsphotonics.5b00180,PhysRevB.91.184207,PhysRevB.86.085146,PhysRevB.91.245128,PhysRevB.91.195147,PhysRevB.92.085107,PhysRevB.77.233104,Tsukerman:11}.
However, very often they required a specific geometry for the metamaterial that motivated the formulation of specific constitutive relations. In contrast, in this work we propose a scheme based on a phenomenological ansatz that can be applied to any periodic metamaterial made of subwavelength, resonant unit-cells to describe homogenized metamaterials.

This paper is structured as follows. In Sec.~\ref{sec:background} we introduce our models from a very general ansatz. Taking a specific fishnet metamaterial as an example, we apply in Sec.~\ref{sec:application} our models to describe its dispersion relation and to show the improvement, compared to a dispersion relation derived with a local materials law. Discussion and conclusion of this work are contained in Sec.~\ref{sec:discussions}.

\section{Constitutive Relations}\label{sec:background}
Let us consider a periodic metamaterial in which the inclusions are intrinsically non-magnetic and reciprocal. We consider time-harmonic fields and only a linear response. It is therefore legitimate to assume that the electromagnetic response may be completely described by a non-local electric response, which can be written as the following integral form \cite{0750626348}
\begin{equation}
\mv{D}(\mv{r},\omega)=  \int_{\mathbb{R}^3}\hat{R}(\mv{r},\mv{r}',\omega)\mv{E}(\mv{r}',\omega)d\mv{r}' \label{eq:relEandD}
\end{equation}
and
\begin{equation}
\mv{H}(\mv{r},\omega)= \mu_0^{-1} \mv{B}(\mv{r},\omega) \,,
\end{equation}
where the kernel $\hat{R}$ represents the electric response tensor, that spatially links in a nonlocal fashion the electric field $\mv{E}$ to the displacement field $\mv{D}$. In a homogeneous medium, the response kernel does not depend on the exact spatial position but only on the difference between two considered points in space. This suggests that the kernel $\hat{R}$ in Eq.~\eqref{eq:relEandD} reduces to a difference kernel, i.e.
\begin{equation}
\mv{D}(\mv{r},\omega)=  \int_{\mathbb{R}^3}\hat{R}(\mv{r}-\mv{r}',\omega)\mv{E}(\mv{r}',\omega)d\mv{r}'\,, \label{eq:convEandD}
\end{equation}
which is a convolution integral. It is more convenient to formulate this constitutive relation in spatial Fourier space, such that the convolution becomes a product
\begin{equation}
\tilde{\mv{D}}(\mv{k},\omega)=  \hat{\tilde{R}}(\mv{k},\omega)\tilde{\mv{E}}(\mv{k},\omega)\,. \label{eq:FTEandD}
\end{equation}
This expression is too general for practical purposes. In order to reach useful constitutive relations, we assume that the unit cells are sub-wavelength and the kernel can be expanded into a Taylor expansion with respect to the spatial frequency $\mv k$. Up to the fourth order expansion, the kernel reads as the following 
\begin{eqnarray}
\tilde{D}_{\alpha}(\mv{k},\omega)&=& 
( \delta_{\alpha \beta}+a_{\alpha \beta}) \tilde{E}_{\beta} \nonumber \\
&+&b_{\alpha\beta\gamma}k_{\gamma}\tilde{E}_{\beta} \nonumber \\
&+&c_{\alpha\beta\gamma\delta}k_{\gamma}k_{\delta}\tilde{E}_{\beta} \nonumber \\
&+&d_{\alpha\beta\gamma\delta\zeta}k_{\gamma}k_{\delta}k_{\zeta}\tilde{E}_{\beta} \nonumber \\
&+&e_{\alpha\beta\gamma\delta\zeta\theta}k_{\gamma}k_{\delta}k_{\zeta}k_{\theta}\tilde{E}_{\beta}  
+ \text{H.O.T.} \,,  \label{eq:constitEandD}
\end{eqnarray}
where Einstein's summation convention has been used. Note that all coefficients are in general complex functions of the frequency $\omega$. 
In the long wavelength limit, i.e.~$|\mv{k}|\rightarrow 0$, spatial dispersion disappears and the constitutive relation \eqref{eq:constitEandD} reduces to the one known from an anisotropic medium with only an electric response in its local description, i.e.
\[\tilde{D}_{\alpha}(\omega) = 
( \delta_{\alpha \beta}+a_{\alpha \beta}) \tilde{E}_{\beta}\,,\]
with the linear electric polarization $\tilde{P}_{\alpha}= a_{\alpha \beta} \tilde{E}_{\beta}$. 
Therefore, the first line of Eq.~\eqref{eq:constitEandD} refers to the local permittivity of an anisotropic medium with
\[
\epsilon_{\alpha\beta}(\omega)= \delta_{\alpha \beta}+{a_{\alpha \beta}(\omega)}\,.
 \]
In order to determine the nature of the higher order terms, we perform an inverse Fourier transform to real space. For plane waves, an inverse Fourier transform to the real space would lead to higher order derivatives of the electric field with respect to spatial coordinates at the same position where the displacement field is to be evaluated, whereupon the constitutive relation becomes
\begin{eqnarray}
D_{\alpha}(\mv{r},\omega)&=& (  \delta_{\alpha \beta}+a_{\alpha \beta} ) E_{\beta} 
+b_{\alpha\beta\gamma}\frac{\partial E_{\beta}}{\partial x_{\gamma} } 
\nonumber \\
&+& c_{\alpha\beta\gamma\delta}   \frac{\partial^2 E_{\beta}}{\partial x_{\gamma}\partial x_{\delta}}
+d_{\alpha\beta\gamma\delta\zeta}\frac{\partial^3 E_{\beta}}{\partial x_{\gamma}\partial x_{\delta}\partial x_{\zeta}}\nonumber \\
&+&e_{\alpha\beta\gamma\delta\zeta\theta}\frac{\partial^4 E_{\beta}}{\partial x_{\gamma}\partial x_{\delta}\partial x_{\zeta}\partial x_{\theta}}
+ \text{H.O.T.} \,,  \label{eq:constitEandDreal}
\end{eqnarray}
where all coefficients were multiplied by a prefactor $(-\mathrm{i})^n$, with $n$ being the order of the spatial derivative.
To be able to practically work with feasible constitutive relation, particular in the context of metamaterials, assumptions to these coefficients have to be made in order to reach well established effective medium theories that shall describe the actual metamaterial. 
 For instance, a frequently made assumption is that 
\begin{align}
b_{\alpha\beta\gamma}(\omega)\frac{\partial}{\partial x_{\gamma}} \overset{!}{=} [\hat{\xi}(\omega)\curl]_{\alpha\beta} \label{eq:assumptionWSD1}
\end{align} 
  and/or 
\begin{align}
 c_{\alpha\beta\gamma\delta}(\omega)\frac{\partial^2}{\partial x_{\gamma}\partial x_{\delta}} \overset{!}{=} [\curl\hat{\gamma}(\omega)\curl]_{\alpha\beta}\,.  \label{eq:assumptionWSD2}
\end{align}  
 If these assumptions are met indeed and it is furthermore assumed that all higher order terms vanish, local constitutive relations identical to those in Eq.~\eqref{eq:LocalBinaisotropy} can be derived by exploiting degrees of freedom that allow to suitably gauge Maxwell's equations. The first assumption leads to a local optical activity (gyrotropy), whereas the second assumption leads to a local magnetic permeability \cite{9056993275}, respectively. We will refer to these assumptions as the weak spatial dispersion (WSD) approximation. ``Spatial dispersion'' because it results form non-local material laws and ``weak" because there is a transformation that gives local constitutive relations, with a local magnetoelectric coupling and a local permeability, hence a local bi-anisotropic medium. However, it has been already shown that these assumptions are not enough to adequately describe the dispersion relation of an actual metamaterial beyond the paraxial regime \cite{PhysRevB.81.035320}. In a principal coordinates system, dispersion relations obtained from WSD are either of the hyperbolic kind, if the principal components of the permittivity or the permeability have opposite signs, or of the elliptic kind otherwise. However, most metamaterials with a nonlocal response usually show dispersion relations that widely differ from hyperbolas or ellipsoids. Exemplarily, we study in the following one example for a metamaterial where the dispersion relation of the fundamental mode indeed differs from that obtained within WSD. This suggests that the WSD approximation is not enough and we, therefore, need to go beyond it and have to retain higher order terms in the expansion. Here, we retain up to the fourth spatial-derivative of the electric field in Eq.~\eqref{eq:constitEandDreal}. 

In order to proceed, we assume WSD as a starting point and extend it into two directions. In our first model, we extend WSD by a, quite similar, but fourth order law in which the fourth order expansion coefficients take the following form
\begin{align}
 e_{\alpha\beta\gamma\delta\zeta\theta}(\omega)\frac{\partial^4 }{\partial x_{\gamma}\partial x_{\delta}\partial x_{\zeta}\partial x_{\theta}} \overset{!}{=}[\curl\curl\hat{\eta}(\omega)\curl\curl]_{\alpha\beta}\,, \label{eq:ellipticform}
\end{align}
where $\hat{\eta}=\text{diag}(\eta_{xx},\eta_{yy},\eta_{zz})$ is a diagonal matrix and represents a higher-order material parameter. This special form is chosen due to mathematical reasons. The coefficients that obey Eq.~\eqref{eq:ellipticform} do not change the variational class of Maxwell's equations and hence, interface conditions may be found by means of variational methods. Even higher order terms again are assumed to vanish.
\\ Our second model remains to be a second order law but it strictly orients on symmetry considerations of a unit-cell of a specific metamaterial. Specifically, it takes more coefficients into account than those that lead to local constitutive relations, i.e.~assumptions \eqref{eq:assumptionWSD1} and \eqref{eq:assumptionWSD2}. The model takes every coefficient into account as required by the symmetry of the considered metamaterial. The tensor elements of the kernel in Eq.~\eqref{eq:FTEandD} reflect the spatial symmetry of the actual structure. The degrees of freedom that the kernel might have can be reduced considerably depending on the symmetry of metamaterials. In this approach, we do not include fourth order terms in expansion \eqref{eq:constitEandD}, as it already gives significant improvements compared to WSD.
The treatment of both approaches simultaneously, i.e.~retaining both unit-cell's symmetry conditions and a fourth order response, in principle, can be applied but the analysis becomes much more involved and is beyond the scope of this paper. In both models and for the sake of simplicity, we only consider metamaterials whose unit cells have a center of symmetry. 
By the presence of this (spatial inversion-)symmetry, all odd terms in expansion \eqref{eq:constitEandDreal} vanish, hence no gyrotropic medium or optical activity is considered. However, it has to be stressed that this is not a general limitation. It is done here to concentrate only on one specific aspect.

 \subsection{Analysis by retaining fourth order response}

\begin{figure}[h]
\centering
\includegraphics[width=1\linewidth]{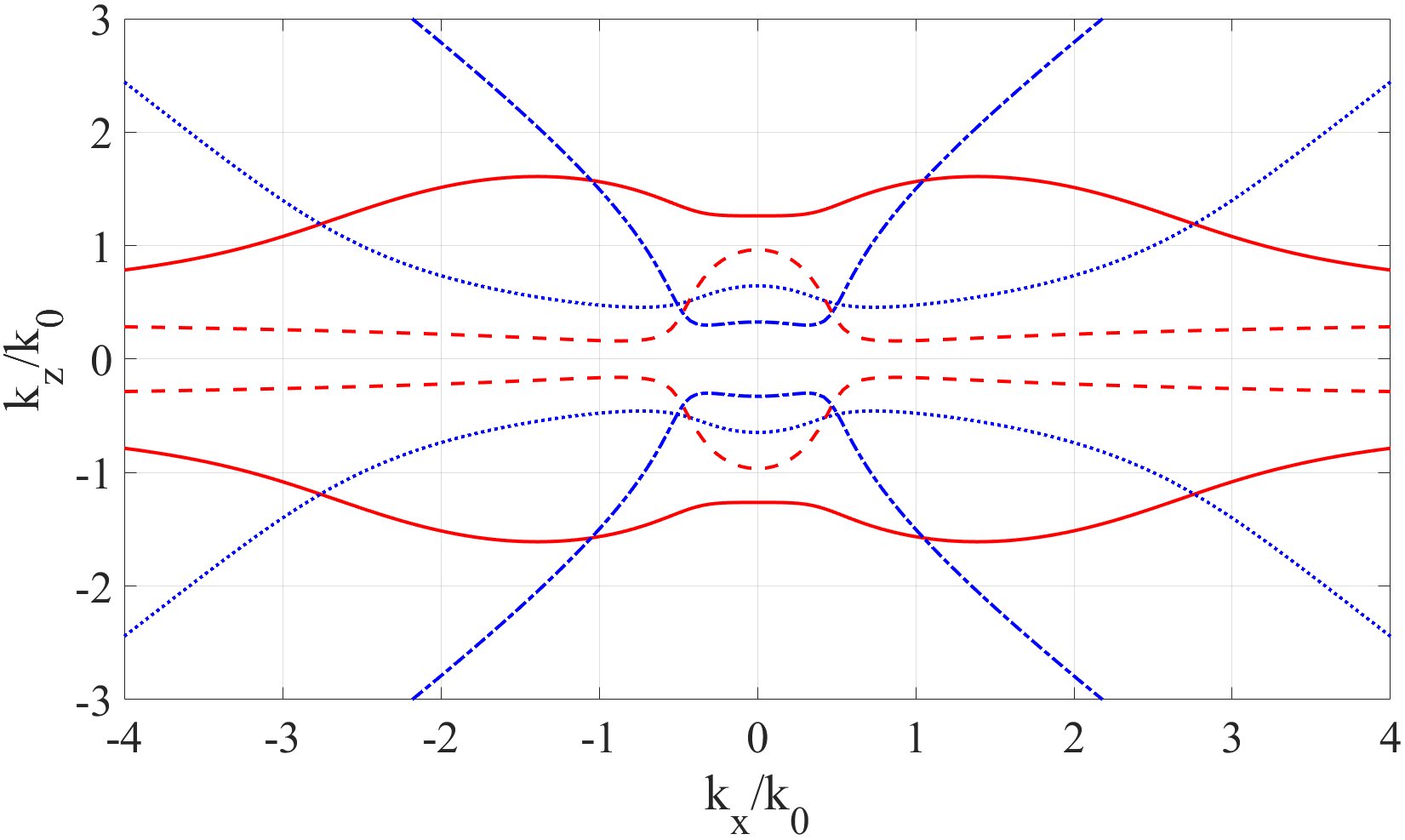} \hfill\vspace{1pt}
\includegraphics[width=1\linewidth]{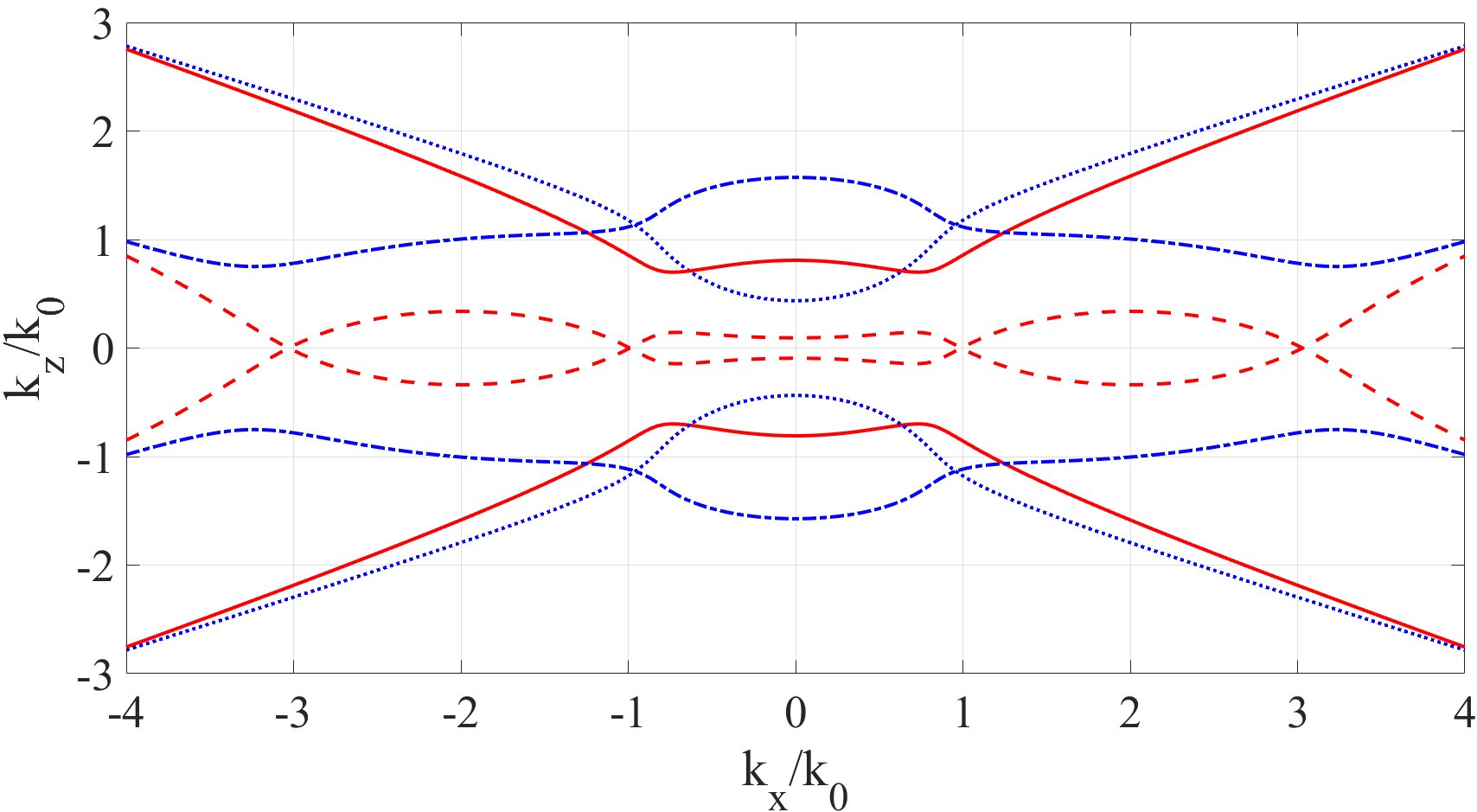} \hfill \vspace{1pt}
\includegraphics[width=1\linewidth]{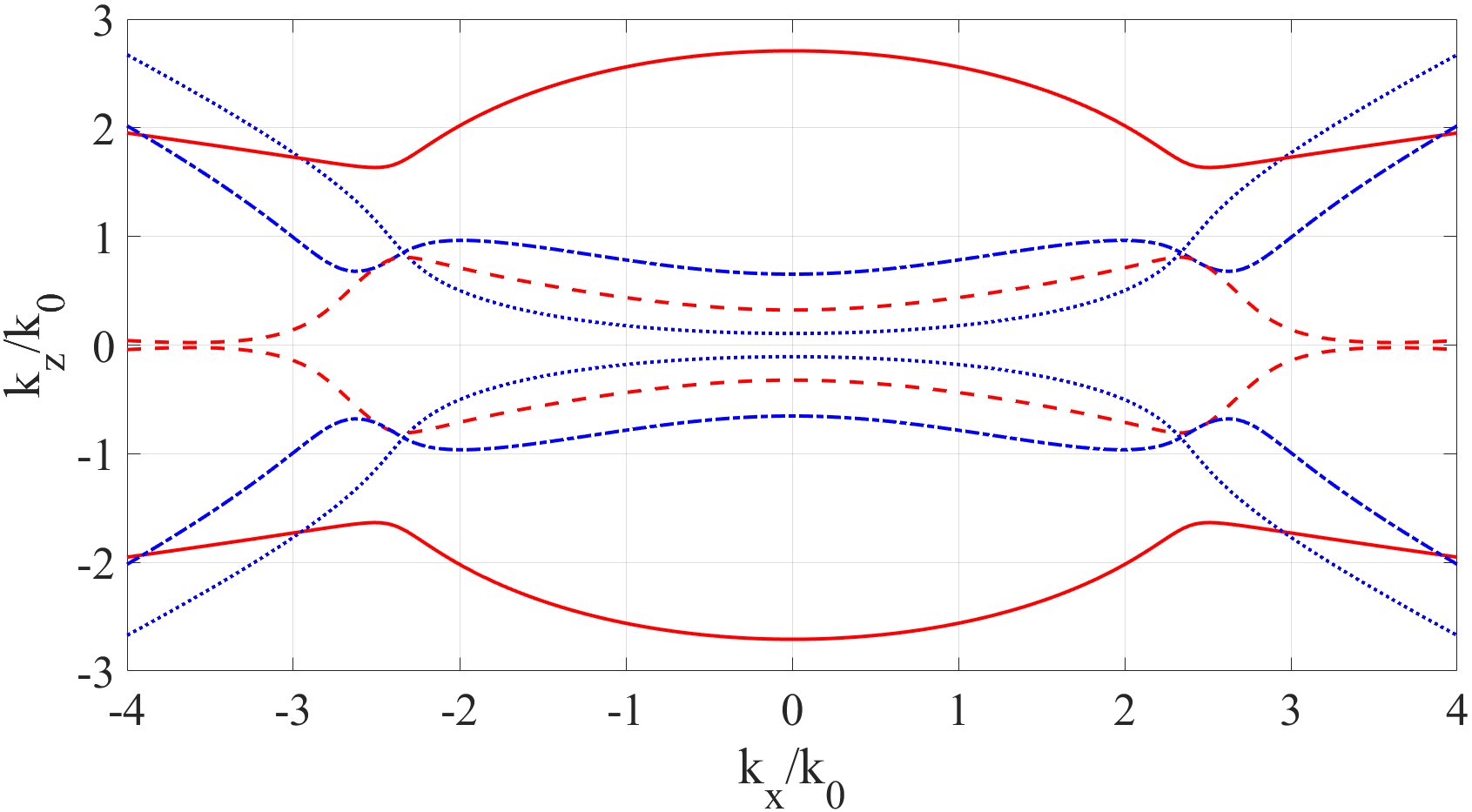} 
\caption{
(Color online) Examples of isofrequency contours decomposed into real (red) and imaginary (blue) parts for the inner $+$ solutions (solid red and dotted blue lines) as well as for the inner $-$ solutions (dashed red and dash-dot blue lines) of Eqs.~ \eqref{eq:Isofreq4thTE} and  \eqref{eq:Isofreq4thTM}. The upper figure shows the isofrequency contours of the TE mode for $p_0^{\text{TE}}= 1+0.5\mathrm{i }\, \mu\mathrm{ m}^{-2}\,, p_1^{\text{TE}}= -1.6-1.5\mathrm{i }\, \mu \mathrm{ m}^{-2}$ and $q_1^{\text{TE}}= 2+0.6\mathrm{i }\, \mu \mathrm{ m}^{-4}$. The center and the bottom figures show the isofrequency contours of the TM modes for $p_0^{\text{TM}}= -1-0.5\mathrm{i }\, \mu\mathrm{ m}^{-2},\, p_1^{\text{TM}}= -3.1+0.1\mathrm{i }\, \mu\mathrm{ m}^{-4},\, q_0^{\text{TM}}= -3.1+0.5\mathrm{i }$ and $q_1^{\text{TM}}= 3+0.5\mathrm{i }$, and $p_0^{\text{TM}}= 3.5-0.5\mathrm{i }\, \mu\mathrm{ m}^{-2},\, p_1^{\text{TM}}= -11.1+0.1\mathrm{i }\, \mu\mathrm{ m}^{-4},\, q_0=^{\text{TM}} -2.1+0.1\mathrm{i }$ and $q_1^{\text{TM}}= 3+0.5\mathrm{i }$, respectively.}
	\label{fig:isofreqfourth}
\end{figure}

Let us now investigate the dispersion relation and the consequences that arise from the following material law
\begin{align}
 \tilde{D}_{\alpha}(\mv{k},\omega) =&\,  \epsilon_{\alpha\alpha}\tilde{E}_{\alpha} -[\mv{k}\times(\hat{\gamma}\mv{k}\times)]_{\alpha\beta}\tilde{E}_{\beta} \nonumber \\
 &+ [\mv{k}\times\mv{k}\times\hat{\eta}\mv{k}\times\mv{k}\times]_{\alpha\beta}\tilde{E}_{\beta}\,.  \label{eq:constitSSD}
\end{align}
This is a direct extension of WSD by the next higher order response such that Maxwell's equations remain of the varuatuinal class. Hence, boundary conditions may be found by means of variational methods.\\
In general, the dispersion relation can be derived by solving the wave-type equation
\begin{equation}
\mv k \times \mv k \times \tilde{\mv E} + k_0^2 \tilde{\mv D}[\tilde{\mv E}]=0\,. \label{eq:waveequation}
\end{equation} 
This represents three coupled equations of second degree for the spatial frequency $\mv k$. Under the assumption that $k_y=0$ and considering the $z$-direction as the principle propagation direction, the dispersion relation for the transversal electric (TE) mode can be found. The dispersion relation expresses here the functional dependency of the frequency and the wave vector components. In TE-polarization the mode has a non-zero electric field component normal to the $k_z$-$k_x$-plane. The dispersion relation reads as
\begin{align*}
k_x^2 \mu_{xx}+k_z^2 \mu_{zz}-\left(k_x^2+k_z^2\right)^2 \eta_{yy}\mu_{xx}\mu_{zz}k_0^2=\epsilon_{yy}\mu_{xx}\mu_{zz}k_0^2 \,,
\end{align*}
where $\mu_{ii}= \frac{1}{1-k_0^2 \gamma_{ii}}$.
The solutions are
\begin{align}\nonumber
k_{z}^2(k_x,k_0) &= -k_x^2+p_0^{\text{TE}}\nonumber \\
&\pm\sqrt{\left(p_0^{\text{TE }}\right)^2-q_1^{\text{TE}}+2k_x^2\left(p_1^{\text{TE}}-p_0^{\text{TE}} \right)} \label{eq:Isofreq4thTE}
\end{align}
with the frequency dependent coefficients
\begin{align}
 q_1^{\text{TE}}(k_0) &= \frac{\epsilon_{yy}(k_0)}{\eta_{yy}(k_0)}\,,  \nonumber \\
p_0^{\text{TE}}(k_0) &= \left[2k_0^2\eta_{yy}(k_0)\mu_{xx}(k_0)\right]^{-1}\,, \nonumber \\
 p_1^{\text{TE}}(k_0) &= \left[2k_0^2\eta_{yy}(k_0)\mu_{zz}(k_0)\right]^{-1} \,,
\label{eq:Coefficients4thTE}
\end{align}
where $q_1^{\text{TE}}$ has the dimension of m$^{-4}$ and both $p_0^{\text{TE}}$ and $p_1^{\text{TE}}$ the dimension of m$^{-2}$.
For the transversal magnetic (TM) mode, i.e.~the mode that has an electric field in the $k_x$-$k_z$-plane, we obtain
\begin{align*}
k_x^2 \epsilon_{xx}+k_z^2\epsilon_{zz}-k_x^2k_z^2(\epsilon_{xx}\eta_{xx}+\epsilon_{zz}\eta_{zz})\mu_{yy}k_0^2 \\
-(k_x^4\epsilon_{xx}\eta_{xx}+k_z^4\epsilon_{zz}\eta_{zz})\mu_{yy}k_0^2=\epsilon_{xx}\epsilon_{zz}\mu_{yy}k_0^2\,,
\end{align*}
with the solutions
\begin{align}\nonumber
k_{z}^2(k_x,k_0) &= -\frac{1}{2}(q_0^{\text{TM}}+q_1^{\text{TM}})k_x^2+p_0^{\text{TM}}\nonumber \\
&\pm\sqrt{\left[p_0^{\text{TM}}+\frac{q_0^{\text{TM}}-q_1^{\text{TM}}}{2}k_x^2\right]^2 -p_1^{\text{TM}}}\,, \label{eq:Isofreq4thTM}
\end{align}
where
\begin{align}
q_{0}^{\text{TM}}(k_0)&=\frac{\epsilon_{xx}(k_0)}{\epsilon_{zz}(k_0)}\,, \nonumber \\
q_{1}^{\text{TM}}(k_0)&=\frac{\eta_{zz}(k_0)}{\eta_{xx}(k_0)} \,, \nonumber \\
p_0^{\text{TM}}(k_0)&=\left[2k_0^{2}\mu_{yy}(k_0)\eta_{xx}(\omega)\right]^{-1}\,, \nonumber \\ 
p_1^{\text{TM}}(k_0)&=\frac{\epsilon_{xx}(k_0)}{\eta_{xx}(k_0)} \,.  \label{eq:Coefficients4th}
\end{align}
Here, both $q_0^{\text{TM}}$ and $q_1^{\text{TM}}$ are dimensionless, while $p_0^{\text{TM}}$ and $p_1^{\text{TM}}$ have the dimensions of m$^{-2}$ and m$^{-4}$, respectively.
Due to the higher number of degrees of freedom, the functional dependency of $k_z(k_x,k_0)$ is more advanced, in comparison to the previously proposed relation \cite{PhysRevB.81.035320},
\begin{align}
k_z^{\text{WSD}^2}(k_x,k_0)= {\alpha_1(k_0)+ \alpha_2 (k_0) k_x^2}\,, \label{eq:IsofreqWSD}
\end{align}
which represents the dispersion relation of a material with local constitutive relations (WSD).
Here, the coefficients $\alpha_1$ and $\alpha_2$ depend on the polarization where for the TE mode they read
\begin{align}
\alpha_1^{\text{TE}}(k_0)&= k_0^2\epsilon_{yy}(k_0)\mu_{xx}(k_0) \,, \nonumber \\
\alpha_2^{\text{TE}}(k_0)&= -\frac{\mu_{xx}(k_0)}{\mu_{zz}(k_0)}\,,
\end{align}
 and
 \begin{align}
\alpha_1^{\text{TM}}(k_0)&= k_0^2\epsilon_{xx}(k_0)\mu_{yy}(k_0) \,, \nonumber \\
\alpha_2^{\text{TM}}(k_0)&=-\frac{\epsilon_{xx}(k_0)}{\epsilon_{zz}(k_0)}\,,
\end{align}
for the TM mode. In general,  isofrequency contours of media described by Eq.~\eqref{eq:IsofreqWSD} are limited to two cases: they are either of hyperbolic or elliptic kind. This limitation is lifted by introducing more complicated dispersion relations, e.g., our fourth order response. To illustrate the possible features an isofrequency contour may have when considering such fourth order constitutive relations, Fig.~\ref{fig:isofreqfourth} shows some plots of $k_z(k_x)$ for a fixed frequency $k_0$ for some generically chosen parameters. The obtained isofrequency contours give rise to more advanced curves. They allow for a homogenous description of an metamaterial with disperive features inaccessible by a local material.

\subsection{Analysis with respect to structure's symmetry}
\begin{figure}[h]
\centering
\includegraphics[width=1\linewidth]{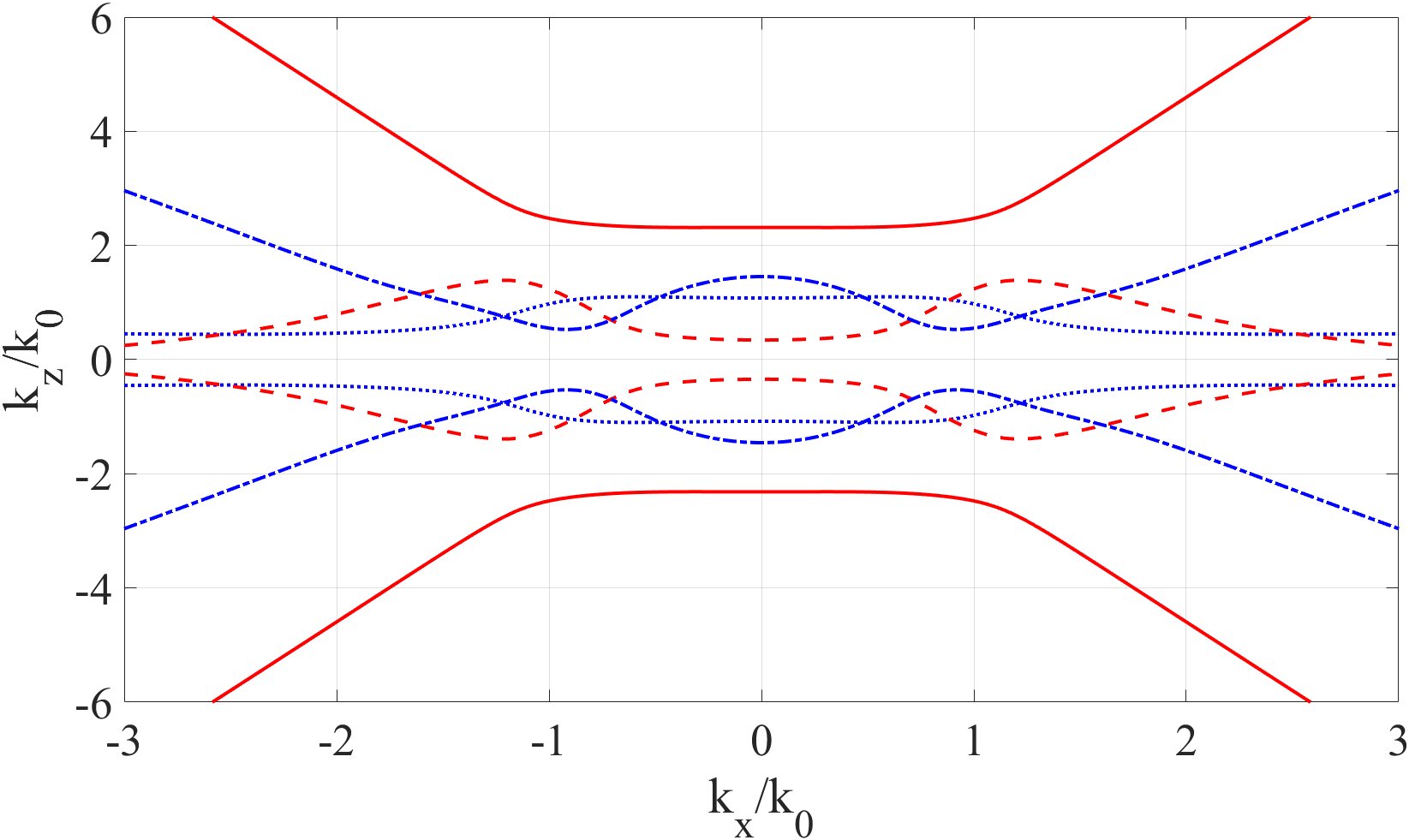}
\hfill	\vspace{1pt}
\includegraphics[width=1\linewidth]{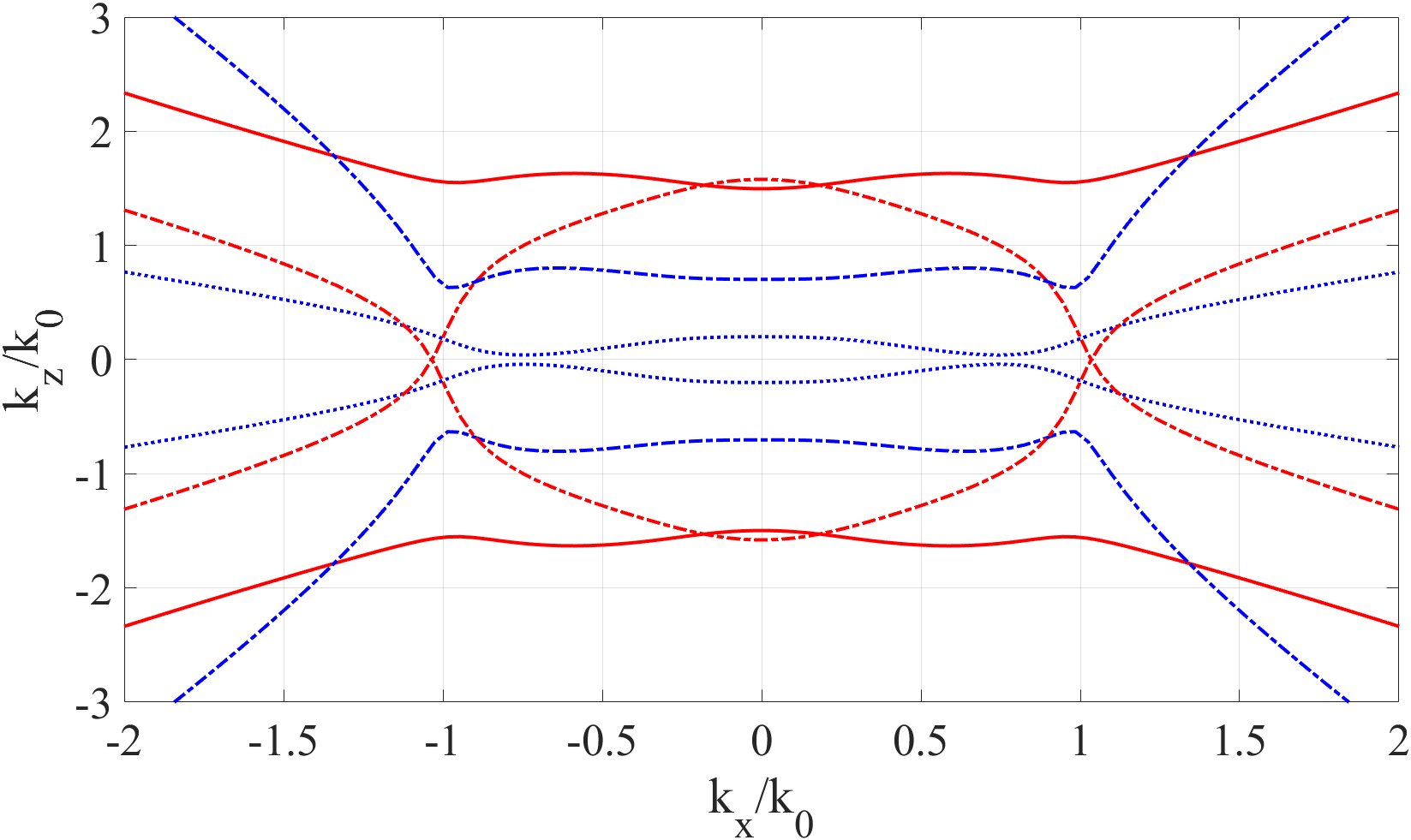}
\caption{(Color online) Examples of isofrequency contours decomposed into real (red) and imaginary (blue) parts for the inner $+$ solutions (solid red and dotted blue lines) as well as for the inner $-$ solutions (dashed red and dash-dot blue lines) of Eq.~\eqref{eq:IsofreqSYM}. The upper figure shows the isofrequency contours for
 $p_0^{\text{TM}}= -1+0.5\mathrm{i }\, \mu\mathrm{ m}^{-2},\, p_1^{\text{TM}}= 2.1+0.1\mathrm{i },\, q_0^{\text{TM}}= 2.1+2.5\mu\mathrm{ m}^{-2}$, and $q_1^{\text{TM}}= -2+0.5\mathrm{i }\,\mu\mathrm{ m}^{-2}$
   while the bottom one for
  $p_0^{\text{TM}}= 1+1.11\mathrm{i }\, \mu\mathrm{ m}^{-2},\, p_1^{\text{TM}}= -1.1+\mathrm{i },\, q_0^{\text{TM}}= 1.1-0.3\mathrm{i }\,\mu\mathrm{ m}^{-2}$, and $q_1^{\text{TM}}= -0.2+2.5\mathrm{i }\,\mu\mathrm{ m}^{-2}$.
   }
	\label{fig:isofreqfsym}
\end{figure}

In this model, instead of taking higher orders in the expansion of the kernel into account, we take a deeper look in the geometry of an actual (real) structure. As an example we consider the fishnet metamaterial  (see Fig.~\ref{fig:fishnet}) as a subject of homogenization. 
The unit cell of the fishnet metamaterial is symmetric under transformation with respect to three orthogonal mirror planes, i.e.~the permittivity distribution obeys 
\[ \epsilon(x,y,z) = \epsilon(\pm x, \pm y, \pm z)\,.
\]
This symmetry class is also known as orthorhombic symmetry \cite{9783662024089}, noted as D$_{2\text{h}}$. According to this consideration, we can write the expansion of the displacement field in the form
\begin{align}
\tilde{D}_{\alpha}(\mv{k},\omega) &= \epsilon_{\alpha\alpha}\tilde{E}_{\alpha} 
 -
[\mv{k}\times(\hat{\gamma}\mv{k}\times)]_{\alpha\beta}\tilde{E}_{\beta} - c_{\alpha\alpha\alpha\alpha}k_{\alpha}^2 \tilde{E}_{\alpha}\,, \nonumber \\  \label{eq:constitD2H}
\end{align}
where Einstein's summation rule applies only to $\beta$. Here, we assume that the local permittivity and $\hat{\gamma}$ are diagonal. More importantly, some of the second order terms may be written in a similar way as in the WSD. The only difference to the WSD is the higher order susceptibility contribution $c_{\alpha\alpha\alpha\alpha}$ that couples $k_{\alpha}^2$ with $E_{\alpha}$. This term, however, cannot be set to zero \textit{a priori} for a metamaterial with the considered symmetry of the fishnet. It has to be taken into account and appears on the same footing as the other second order terms. These other terms in $k^2$ are assumed to obey the condition that yields a local magnetic response, i.e.~Eq.~\eqref{eq:assumptionWSD2}. The dispersion relation follows from solving the wave equation as described in Eq.~\eqref{eq:waveequation}. We assume equally that $k_y=0$ and we study a plane with a wavevector in the $k_x$-$k_z$-plane. The dispersion relation for the TE mode is then
\begin{align}
& k_z^2\mu_{zz} 
+ k_x^2\mu_{xx} = \mu_{xx}\mu_{zz}\epsilon_{yy}k_0^2\,, \label{eq:DispSymTE}
\end{align}
with the solutions
\begin{align}
k_{z}^2 (k_x,k_0)= { p_0^{\text{TE}}+q_0^{\text{TE}} k_x^2 }
\end{align}
where
\begin{align}
q_0^{\text{TE}}(k_0) &= -\frac{\mu_{xx}}{\mu_{zz}}\,, \\
 p_0^{\text{TE}}(k_0) &= \epsilon_{yy}\mu_{xx} k_0^2\,.
\end{align}
The fact that the TE mode does not experience any strong spatial dispersion relies on the nature of the expansion in Eq.~\eqref{eq:constitD2H} with the nonlocal response that couples only in the direction of the electric field, hence no cross coupling between the displacement field and the electric field as can be seen in Eq.~\eqref{eq:constitD2H}. In contrast, a more complicated dispersion relation is found  for the TM modes with
\begin{align}
& k_z^2(\epsilon_{zz}+c_Z \epsilon_{xx}\mu_{yy}k_0^2)  \nonumber
+ k_x^2(\epsilon_{xx}+c_X \epsilon_{zz}\mu_{yy}k_0^2) \\
-& k_z^2 k_x^2 c_Z c_X \mu_{yy}k_0^2 -k_z^4c_Z- k_x^4 c_X = \epsilon_{xx}\epsilon_{zz}\mu_{yy}k_0^2\,, \label{eq:DispSymTM}
\end{align}
where $\mu_{ii}= \frac{1}{1-k_0^2\gamma_{ii}}$ and $c_J=c_{jjjj}$. This equation is biquadratic in $k_x$ and $k_z$ and describes more complicated isofrequencies than quadratic equations, e.g.~given by WSD. Of course, the limit $(c_X,c_Z) \rightarrow (0,0)$ restores the dispersion relation given by WSD. The solutions of Eq.~\eqref{eq:DispSymTM} are
\begin{align}
k_{z}^2(k_x,k_0) &= p_1^{\text{TM}} k_x^2+q_0^{\text{TM}}+p_0^{\text{TM}} \nonumber \\  &\pm \sqrt{(p_1^{\text{TM}}k_x^2+p_0^{\text{TM}}-q_0^{\text{TM}})^2+2q_1^{\text{TM}}\left(\frac{p_1^{\text{TM}}}{p_0^{\text{TM}}}k_x^4+k_x^2\right)} \label{eq:IsofreqSYM}
\end{align}
with the related coefficients that read
\begin{align}
q_0^{\text{TM}}(k_0) &= \frac{\epsilon_{zz}(k_0)}{2c_Z(k_0)}\,, \nonumber \\
 p_0^{\text{TM}}(k_0) &=  \frac{k_0^2}{2}\epsilon_{xx}(k_0)\mu_{yy}(k_0)\,, \nonumber \\
q_1^{\text{TM}}(k_0)&= \frac{\epsilon_{xx}(k_0)}{2c_Z(k_0)}\,, \nonumber \\  
 p_1^{\text{TM}}(k_0) &=  -\frac{k_0^2}{2}c_X(k_0)\mu_{yy}(k_0)\,. \label{eq:CoefficientsSYM}
\end{align}
It has to be noted that all coefficients have the dimension of m$^{-2}$, except $p_1$ being dimensionless.
Here, we have four independent coefficients which increase the number of degrees of freedom to four. Previously, e.g.~in Eq.~\eqref{eq:IsofreqWSD}, the dispersion relation contained only two independent coefficients, $\alpha_1$ and $\alpha_2$, hence only two degrees of freedom. Moreover, and this holds for both Strong Spatial Dispersion (SSD) models, each of the Eqs.~(\ref{eq:IsofreqSYM}, \ref{eq:Isofreq4thTE}, \ref{eq:Isofreq4thTM})  yield four possible solutions for $k_z(k_x,k_0)$ from which two with positive and two with negative imaginary parts. We consider here only solutions with a positive imaginary part as they describe exponentially damped solutions in our principle propagation direction. In contrast the two solutions with a negative imaginary part would correspond to exponentially growing solutions. They are unphysical for a passive medium. However, this still suggests that in the actual homogeneous medium characterized by material laws beyond WSD, more than a single mode is excited at the interface where the continuity of the tangential wave vector components dictates which modes are excited. To simplify the analysis, we concentrate on investigating the fundamental mode, i.e.~the mode with the smallest positive imaginary part. In general, the imaginary parts of the two solutions with $\Im(k_z)>0$ may cross and a mode transition has to be taken into account.
Here as well, Fig.~\ref{fig:isofreqfsym} shows some isofrequency contours from selected parameter sets. The complexity of their shapes suggests the ability to capture the effects of strong spatial dispersion. 

In the next section, we show the importance of retaining these nonlocal effects in the effective description of the metamaterial by directly showing the improvements that follow from taking SSD into account, as introduced in Eqs.~\eqref{eq:constitSSD}  and \eqref{eq:constitD2H}.

   \section{Application to a fishnet metamaterial}\label{sec:application}
   \begin{figure}[h]
\centering
\includegraphics[width=0.710\linewidth]{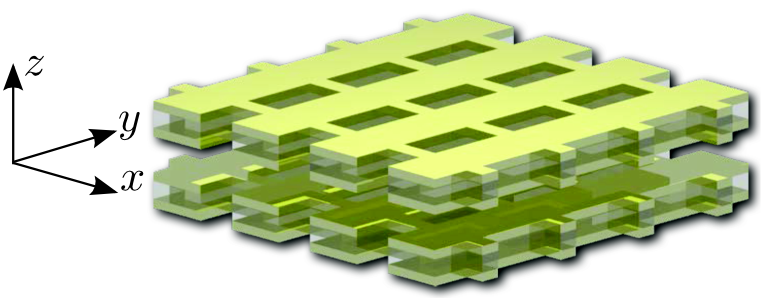}
\caption{Fishnet metamaterial consisting of a bi-periodic structure with periods $\Lambda_x=\Lambda_y=600\,$nm and $\Lambda_z=200\,$nm with rectangular holes with the width $w_x= 100\,$nm and $w_y= 316\,$nm. It compromises a stack of layers made of two $45\,$nm Ag layers separated by a thin dielectric spacer, $30\,$nm of MgF$_2$. The remaining space is filled with air.}
	\label{fig:fishnet}
\end{figure}
  \begin{figure}[h]
\centering
\includegraphics[width=0.5\linewidth]{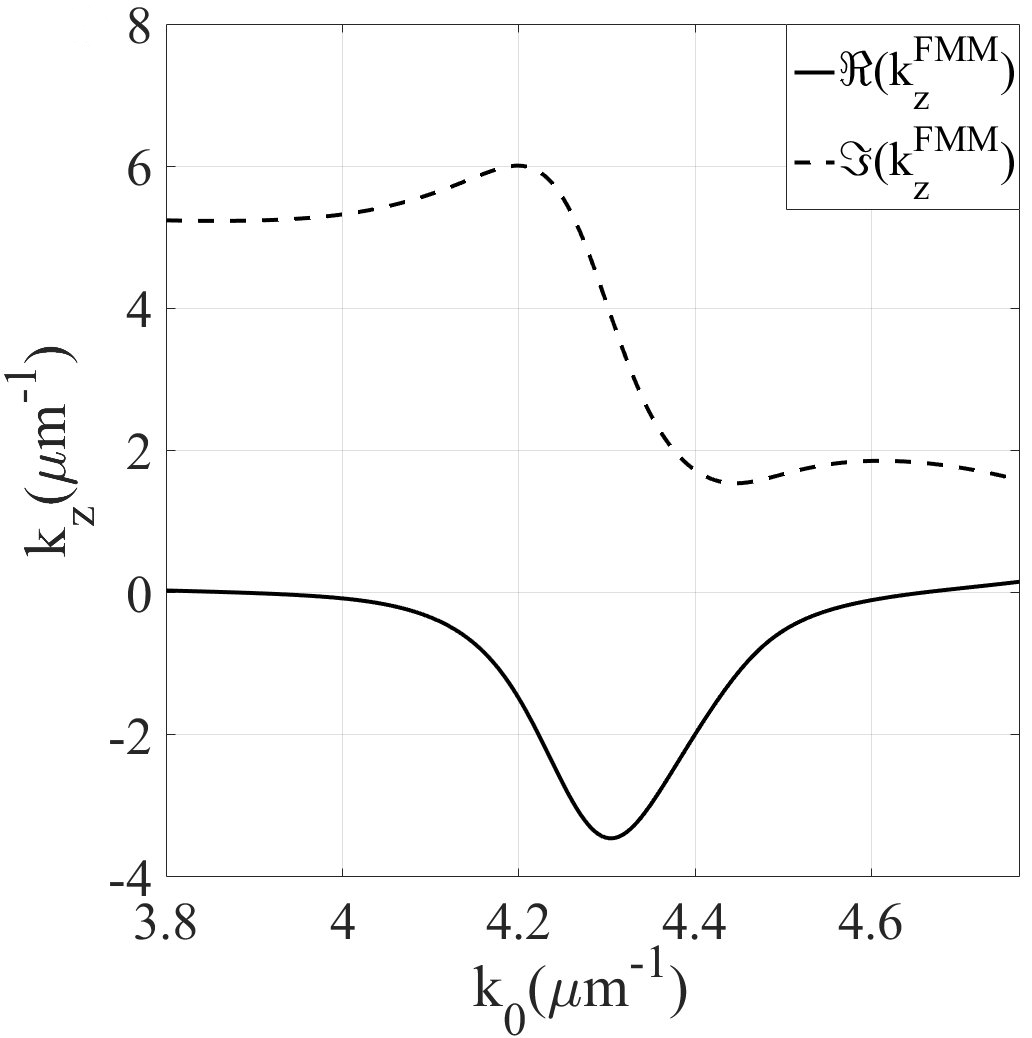}\includegraphics[width=0.5\linewidth]{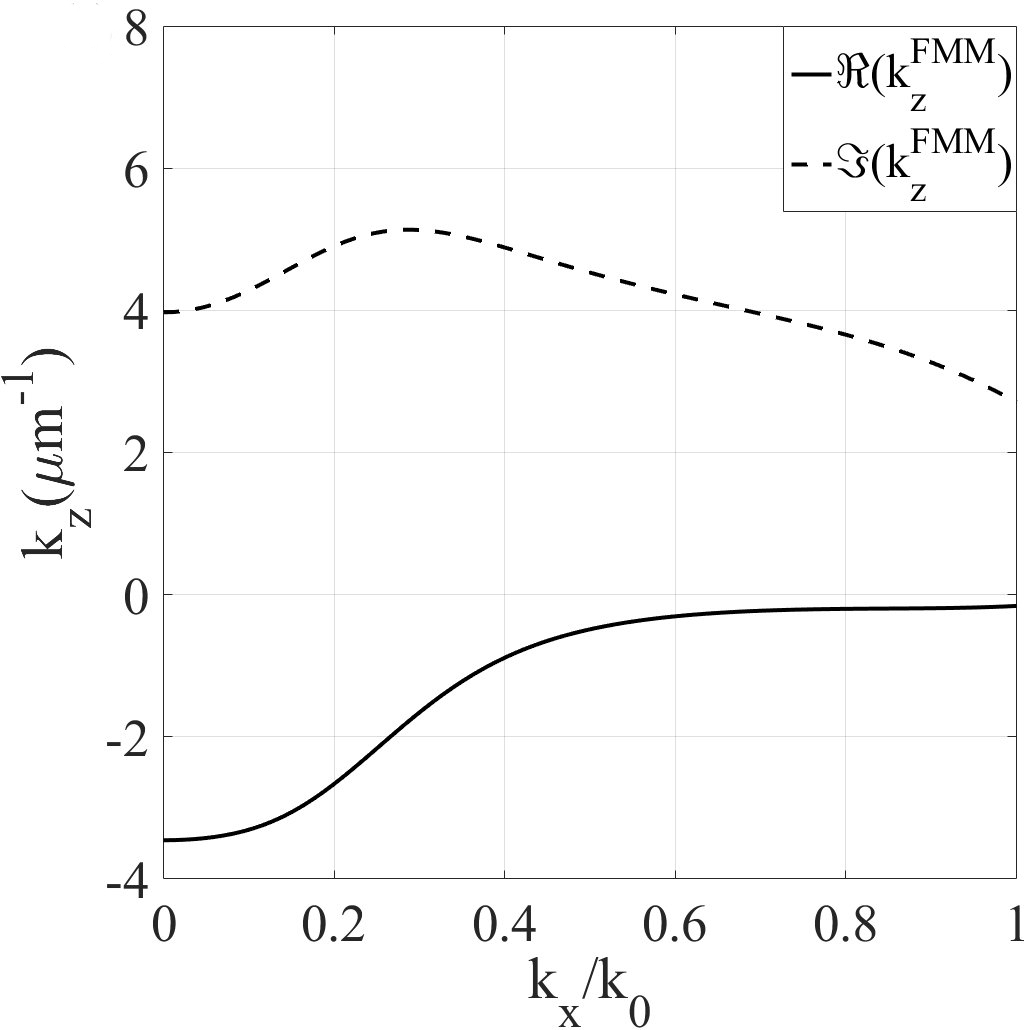}
\caption{(left) Dispersion relation of the fundamental TM mode for $k_x=k_y=0$ calculated by a plane wave expansion ansatz using the Fourier modal method (FMM) algorithm. It shows a resonance around $k_0= 4.3~\mu$m$^{-1}$ in which $\Re (k_z)<0$. (right) Isofrequency contour in the $xz$-plane at the resonance wavenumber.}
	\label{fig:fishnetdispersion}
\end{figure}

In this section, a numerical experiment is done to get access to the dispersion relation of an actual structure as a reference. We consider the fishnet metamaterial shown in Fig.~\ref{fig:fishnet} as an example, which is known to exhibit a negative refractive index at optical frequencies. The geometrical parameters are taken from literature \cite{Dolling06}. The unit cell's dimensions are $\Lambda_x=\Lambda_y=600$ nm, $\Lambda_z=200$ nm. The rectangular holes are made of perpendicularly aligned nanowires with thicknesses of $w_x= 100\,$nm and $w_y= 316\,$nm. It is made of two $45\,$nm Silver (Ag) layers where their permittivity obey the Drude model for metals, with the plasma frequency being $\omega_p =13700 \text{ THz}$ and the relaxation rate $\Gamma = 85 \text{ THz}$. These layers are separated by a $30\,$nm Magnesium fluoride (MgF$_2$) spacer whose permittivity is assumed to be nondispersive ($\epsilon_{\text{MgF}_2}=1.9044$) in the frequency range of interest. Furthermore, the unit cell is symmetric with respect to spatial inversion, i.e.~$\epsilon(\mv{r})=\epsilon(-\mv{r})$. 

\begin{figure}[h]
\centering
\includegraphics[width=1\linewidth]{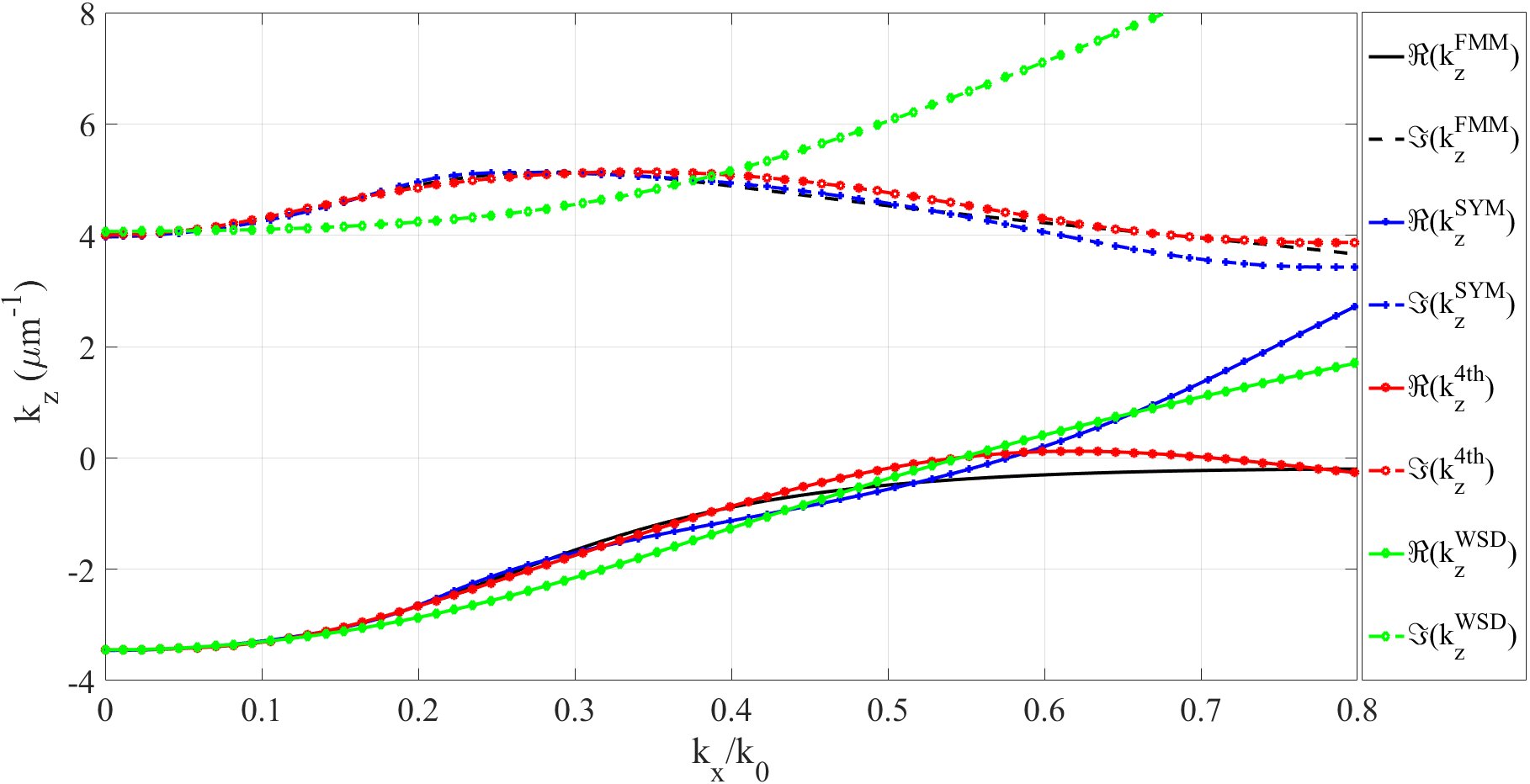}
\caption{(colors online) Isofrequency contours $k_z=k_z(k_x)$ at the resonance frequency of the fishnet corresponding to a wavenumber of $k_0= 4.3\,\mu$m$^{-1}$. Solid (dashed) curve represent real (imaginary) part. The blue (crosses) curves are obtained from fitting Eq.~\eqref{eq:IsofreqSYM} to the reference curve (black). It shows a good agreement up to $k_x=0.3k_0$. 
The red (bullets) curves are obtained from fitting Eq.~\eqref{eq:Isofreq4thTM} to the reference curve and shows a good agreement up to $k_x=0.4k_0$. Meanwhile, the green (diamonds) curves, which are obtained from WSD, are showing only an agreement within the paraxial regime, i.e.~for $k_x < 0.1{k_0}$.
}\label{fig:isofreqresults}
\end{figure}

In order to calculate the Bloch mode dispersion relation, i.e.~$k_z=k_z(k_x,k_y,k_0)$ where $k_0$ represents the wave number in free space, a plane-wave expansion ansatz is numerically performed. In general, $k_z$ can be complex. Its real part, $\Re(k_z)$, refers to the oscillatory part while its imaginary part, $\Im(k_z)$, denotes the energy loss in the principal propagation direction. We restrict our considerations to the solutions with $\Im(k_z)>0$. The dominating Bloch mode, i.e.~the fundamental Bloch mode that prevails after a finite propagation length is the one with the smallest positive $\Im(k_z)$ as all higher modes experience much stronger damping. Fig.~\ref{fig:fishnetdispersion} shows both dispersion relation and isofrequency contours of the fundamental Bloch mode for different transverse vectors $k_x \geq0$ at a fixed frequency and for different frequencies at a fixed transverse wave vector, respectively. Here we restrict ourselves to the $k_x$-$k_z$-plane, i.e.~$k_y=0$. The mode is TM polarized. For wavenumbers around $4.3\, \mu$m$^{-1}$, we observe negative $\Re (k_z)$, which implies a negative index, i.e.~momentum and energy flux propagate in opposite directions. We are basically interested in homogenizing the material around this resonance wavenumber. Figure \ref{fig:fishnetdispersion} (right) shows the isofrequency contour for $k_0=4.3\, \mu$m$^{-1}$ of the real structure. This numerically obtained isofrequency contour of the actual metamaterial has to be reproduced by the dispersion relation of the effective medium from WSD as well as from both SSD with some fixed set of parameters. The comparison is based on a least absolute deviations fit by optimizing the parameters $(q_0,p_0,q_1,p_1)$ of fundamental TM modes from the SSD models and $(\alpha_1,\alpha_2)$ from the WSD. As a quantity of measure, we define the merit function as the following
\begin{align}
 \delta(k_0)= \frac{\sum_{k_x} \big|1-\frac{{k_z^{\text{i}}}^2(k_x,k_0)}{{k_z^{\text{FMM}}}^2(k_x,k_0)}\big| w(k_x)}{\sum_{k_x}  w(k_x)}\,, \label{eq:meritfunction}
\end{align}
with a suitably defined weight function $w(k_x)$. Here $\text{i} \in \{\text{WSD, 4th, SYM}\}$ denotes the model taken into account. 4th and SYM refer to the fourth-order and to the symmetry model, respectively.
As all the expressions were derived from a Taylor expansion for small $\mv k$ (see Eq.~\eqref{eq:constitEandD}), it is legitimate to introduce a weight $w(k_x)$ such that the fitting procedure is more focused for small $k_x$. Here we chose an exponentially decreasing dependency, i.e.
\begin{align}
w(k_x)=\text{e}^{-\alpha k_x}\,,
\end{align}
where $\alpha$ was chosen to be $\alpha = 2.5 \Lambda_x$, with $\Lambda_x=0.6\, \mu$m being the lateral period of the fishnet structure.
The results for a selected frequency -- here we chose the worst case scenario, i.e.~at the resonance -- when considering the optimized parameters are depicted in Fig.~\ref{fig:isofreqresults}. It shows the isofrequency contour in both real and imaginary part of the dispersion relation numerically calculated for the actual fishnet metamaterial and the dispersion relation obtained for the best fit of parameters at the resonance frequency for the three different models considered. The parameter set of the SSD models for the best fit and the right signs are summarized in Table \ref{tab:bestfitparameters}. For the parameters of WSD, i.e relation \eqref{eq:IsofreqWSD}, we obtain  $\alpha_1^{\text{TM}}= (-5.85 -28.12\mathrm{i})\mu$m$^{-2}$ and $\alpha_2^{\text{TM}}=-2.49 + 3.79\mathrm{i}$. Clearly, WSD is only in a good agreement with the isofrequency contour of the real structure  in the paraxial regime, i.e.~$k_x \ll k_0$. Beyond the paraxial regime, we recognize from the shape of the black curves of $k_z(k_x)$, that WSD, which by nature is either an ellipse or a hyperbola, is not enough to describe the dispersion relation of such complicated structure. This limitation can be lifted by considering nonlocal constitutive relations as proposed above. The actual dispersion relation can be much better described when homogenizing the metamaterial with constitutive relations beyond the local model.
To quantify the actual improvement, we study the merit function as a function of the frequency for the three different constitutive relations obtained in here. Each value for the merit function has been obtained from an individual fit at a specific frequency.


\begin{center}
\begin{table}
\begin{tabular}{ |c|c|c| } 
 \hline
  Model & & \\ and sign: & 4th(-) & SYM(+) 
  \\ 
  \hline \hline
  \rule{0pt}{3ex}    
  $p_0^{\text{TM}}$ & $(-1.01 -56.62\mathrm{i})\mu$m$^{-2}$ & $(-1.97 -13.85\mathrm{i})\mu$m$^{-2}$ \\ 
  \rule{0pt}{3ex}    
 $p_1^{\text{TM}}$ & $(-2.35 + 0.28\mathrm{i})\cdot 10^3\mu$m$^{-2}$ & $-7.21 - 1.12\mathrm{i}$ \\ 
  \rule{0pt}{3ex}    
 $q_0^{\text{TM}}$ & $0.8 04+ 0.807\mathrm{i}$ & $(-18.2 -10.61\mathrm{i})\mu$m$^{-2}$ \\ 
  \rule{0pt}{3ex}    
 $q_1^{\text{TM}}$ & $-26.43 -21.80\mathrm{i}$ & $(-27.7 +33.95\mathrm{i})\mu$m$^{-2}$ \\ 
 \hline
\end{tabular} 
\caption{Parameter set of the SSD models for the best fit as shown in Fig.~\ref{fig:isofreqresults}. Depending on the model there is always one sign (the $\pm$ sign in Eqs.~\eqref{eq:Isofreq4thTM} and \eqref{eq:IsofreqSYM}) that fits better to the fundamental Bloch mode. Bear in mind that parameters from different models have different expressions and dimensions.}
\label{tab:bestfitparameters}
\end{table}
\end{center}

\begin{figure}[h]
\centering
\includegraphics[width=1\linewidth]{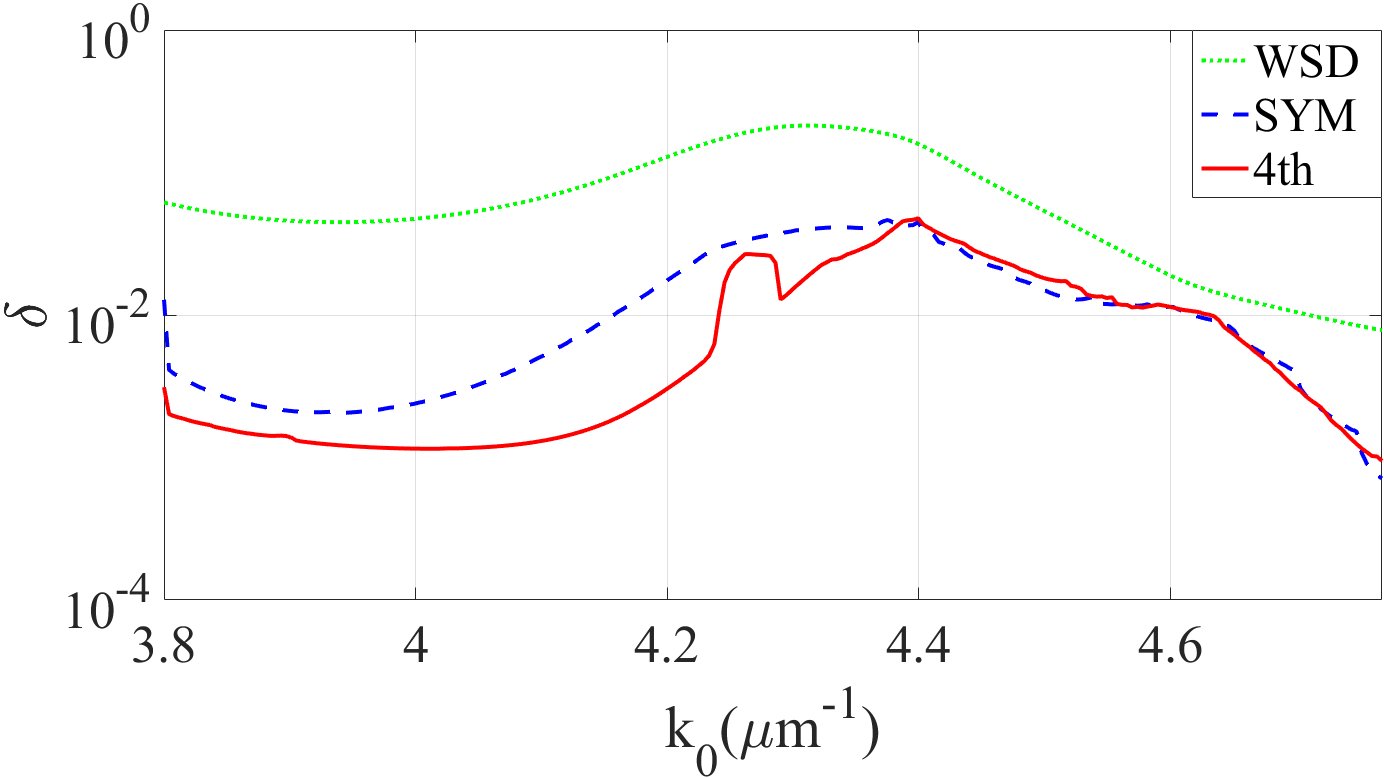}
\caption{
Deviations from reference (FMM) in logarithmic scale. Over all frequencies, modeling metamaterials with nonlocal material laws makes more sense for a realistic homogenization.
} \label{fig:meritfunction}
\end{figure}

Figure \ref{fig:meritfunction} shows the improvements in effectively describing the metamaterial with nonlocal constitutive relations for all of the simulated frequencies. Both SSD models are more accurate than WSD. The integrated error that expresses how good a specific constitutive relation in a homogenized medium can explain the actual dispersion relation of the given metamaterial, is in average two orders of magnitude better for the nonlocal material laws. In resonance, the deviation is strongest irrespective of the considered model. However, this is somehow expected that the effective description tends to be inaccurate in the resonance regime. Nevertheless, the findings immediately implies retaining nonlocal constitutive relations is required for a more realistic homogenization of metamaterials. 
 
\section{Discussion and Conclusions}\label{sec:discussions}
The aim of this paper was to introduce on the one hand a viable route to describe homogenized metamaterial beyond the assumption of a local response and to show the importance of retaining nonlocal material laws in order to adequately describe the dispersion relation of a homogenized metamaterial. Of course we are not the first to work in this direction, but the approach we have choosen here is generally applicable and does not hinge on the assumption of a specific geometry of the metamaterial. At the example of a fishnet metamaterial we have shown that WSD is not enough to properly capture the dispersion relation as their functional dependency and their isofrequency contours -- either hyperbolic or elliptic -- are too simplistic to give accurate predictions beyond the paraxial regime. Significant improvement only comes by introducing nonlocal material laws in their effective description. These come with more degrees of freedom, hence more complicated isofrequency come into play and give a better description of metamaterials. In our work we have studied the light propagation in the bulk of a metamaterial and we obtained all the coefficients that enter the dispersion relation by means of optimization in comparision to the numerically calculated dispersion relation of an actual metamaterial. However, for a complete inversion, i.e.~to obtain the individual material parameters, requires finding the interface conditions between one ordinary medium and one with nonlocal material laws. This is subject of ongoing research.

\section*{Acknowledgements}
We gratefully acknowledge financial support by the Deutsche Forschungsgemeinschaft (DFG) through CRC 1173. K.M. also acknowledges support from the Karlsruhe School of Optics and Photonics (KSOP). We would like to thank Radius Suryadharma and Andreas Vetter for proofreading the manuscript.
\bibliography{paperbib}

\end{document}